\documentclass[nologo,11pt,a4paper]{ETHpaper}
\usepackage{hyperref}
\usepackage{amsmath, amssymb}
\usepackage{graphicx}
\usepackage{subfigure}
\usepackage{scalefnt}
\usepackage[numbers,sort&compress]{natbib} 
\usepackage{enumitem}

\title{Reaction-Diffusion Processes on Interconnected\\ Scale-Free Networks} 
\titlealternative{Reaction-Diffusion Processes on Interconnected Scale-Free Networks\\
Journal Ref: Phys. Rev. E {\bf 92}, 020801(R), 2015}
\author{Antonios Garas} 
\authoralternative{Antonios Garas}
\address{Chair of Systems Design, ETH Zurich, Switzerland\\
  \url{www.sg.ethz.ch}}
\www{\url{http://www.sg.ethz.ch}}
\makeframing

\begin{document}

\maketitle

\begin{abstract}
  We study the two particle annihilation reaction $A+B\rightarrow \emptyset$ on interconnected scale free networks, using
  different interconnecting strategies. We explore how the mixing of particles and the process evolution are influenced by
  the number of interconnecting links, by their functional properties, and by the interconnectivity strategies in use. 
  We show that the reaction rates on this system are faster than what was observed in
  other topologies, 
  due to the
  better particle mixing which suppresses the segregation effect, inline with previous studies performed on single scale free networks.
\end{abstract}

Using complex networks to describe real systems
is becoming standard practice~\cite{newman2010,cohen2010}, but, understanding how networks interact
in an increasingly interconnected world~\cite{Garas2015} is an open challenge for network science.  
Networks in general retain their identity despite the existence of interconnections, for
example a communication network does not change its role when it is connected to a power network. This makes our knowledge
about isolated networks relevant for the interconnected case. However, there exist network properties that are strongly affected by
interconnectivity~\cite{Bashan2013,Gao2011,Gomez2013,Baxter2012a,Wang2013a,Cardillo2013,Brummitt2012a,Aguirre2014}

In addition, interconnecting links may be different (with respect to their function) than the normal links within the networks. In
the example at hand, via interconnecting links we provide power to the communication network, and we control how the load is
distributed in the power network. Thus, failure of these links has severe consequences, which is why in this case they are called
dependency links~\cite{Buldyrev2010}. In such cases extra care should be taken, and interconnected networks should not be studied
as isolated networks with distinct communities. Indeed, recently it was shown that an interconnected system of networks may be
either in a regime where the various networks are structurally independent, or in a regime where they are strongly coupled and the
system behaves like one large network~\cite{Radicchi2013,martin2014algebraic}. 
This can influence at large the evolution of dynamical processes on such systems.

In this work we provide a detailed numerical study of how interconnectivity affects the evolution of a diffusion-reaction
dynamical process, when the reaction evolves on an interconnected network substrate. More precisely, we study the reaction rates
of the annihilation reaction $A+B\rightarrow \emptyset$ on coupled Scale Free Networks (SFN) using different interconnecting
strategies.

In SFN the probability to find a node with $k$ connections (degree) is given by $P(k)\sim k^{-\gamma}$. Such degree distributions
allow the existence of a small number of nodes with very large number of links (i.e. hubs), while the majority of the nodes have
only a few links. The number of hubs depends on the exponent $\gamma$, which typically has values in
the range $2 < \gamma < 4$. Small values of $\gamma$ lead to heterogeneous networks, while as $\gamma$ increases, especially when
$\gamma > 3$, the networks are getting more homogeneous.

The annihilation reaction $A+B\rightarrow \emptyset$, is an exemplary case of diffusion-controlled reactions which were used to
model chemical reactions, epidemics, and other dynamical processes~\cite{Van1992,Murray2004}.  In general, the quantity of
interest is the concentration of particles $\rho(t)$ that remain in the system at a given time $t$, which follows
\begin{equation}
 \frac{1}{\rho(t)}-\frac{1}{\rho_0}=\kappa t^{f},
 \label{eq:cons}
\end{equation}
where $\rho_0$ is the initial particle concentration, $\kappa$ is the rate constant, and $f$ is the exponent that determines the reaction rate. The maximum value of $f$ for various
topologies is set by the mean-field asymptotic limit to $f=1$, while the non-classical kinetics predicts that the exponent $f$
depends on the dimensionality $d$ of the space where the process evolves as $f=d/d_{c}$ for $d\leq d_{c}$, and $f=1$ for
$d>d_{c}$~\cite{Ovchinnikov1978,Toussaint1983,Torney1983,Havlin1987,Krapivsky2010}. Here, $d_{c}$ is the upper critical dimension,
which for the $A+B$ reaction is $d_{c}=4$. Surprisingly, however, it was shown that when SFN are used as substrate $f$ can obtain
values larger than one~\cite{Gallos2004} and particles do not segregate, similar to what was observed in systems with Levy
mixing~\cite{Zumofen1996}.

\begin{figure*}[t]
\includegraphics[width=0.22\columnwidth]{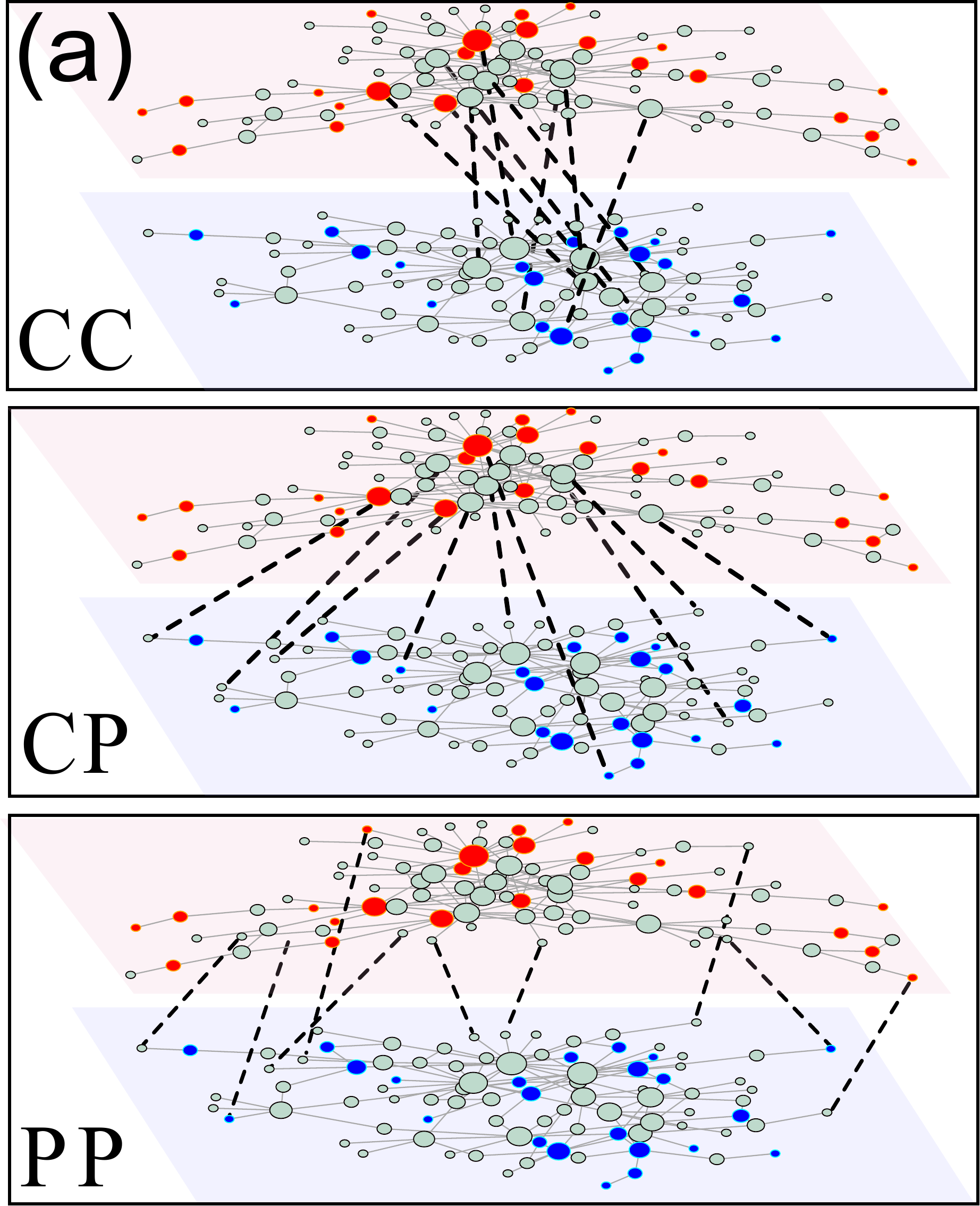}
\hfill
\includegraphics[width=0.24\columnwidth]{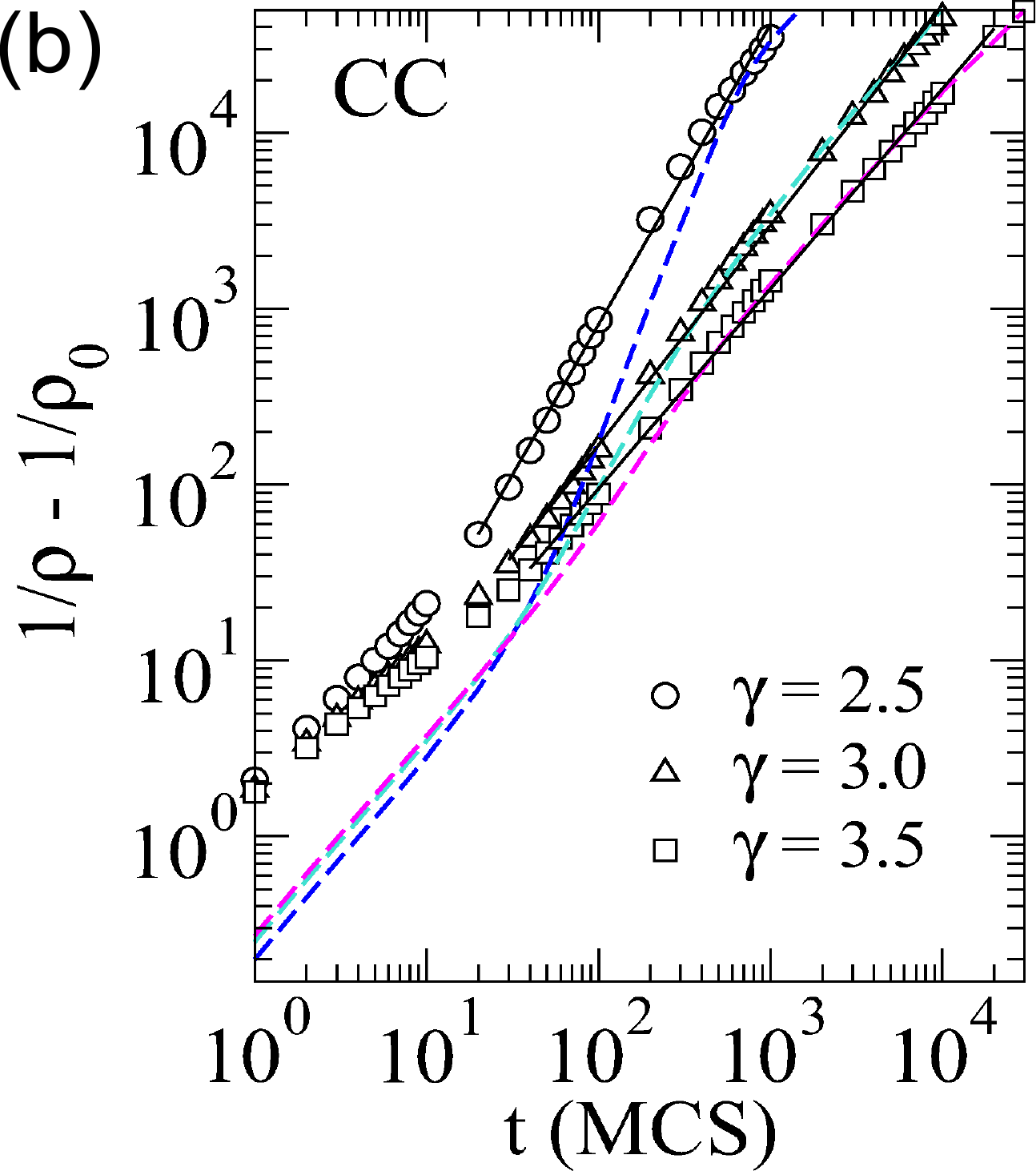}
\hfill
\includegraphics[width=0.24\columnwidth]{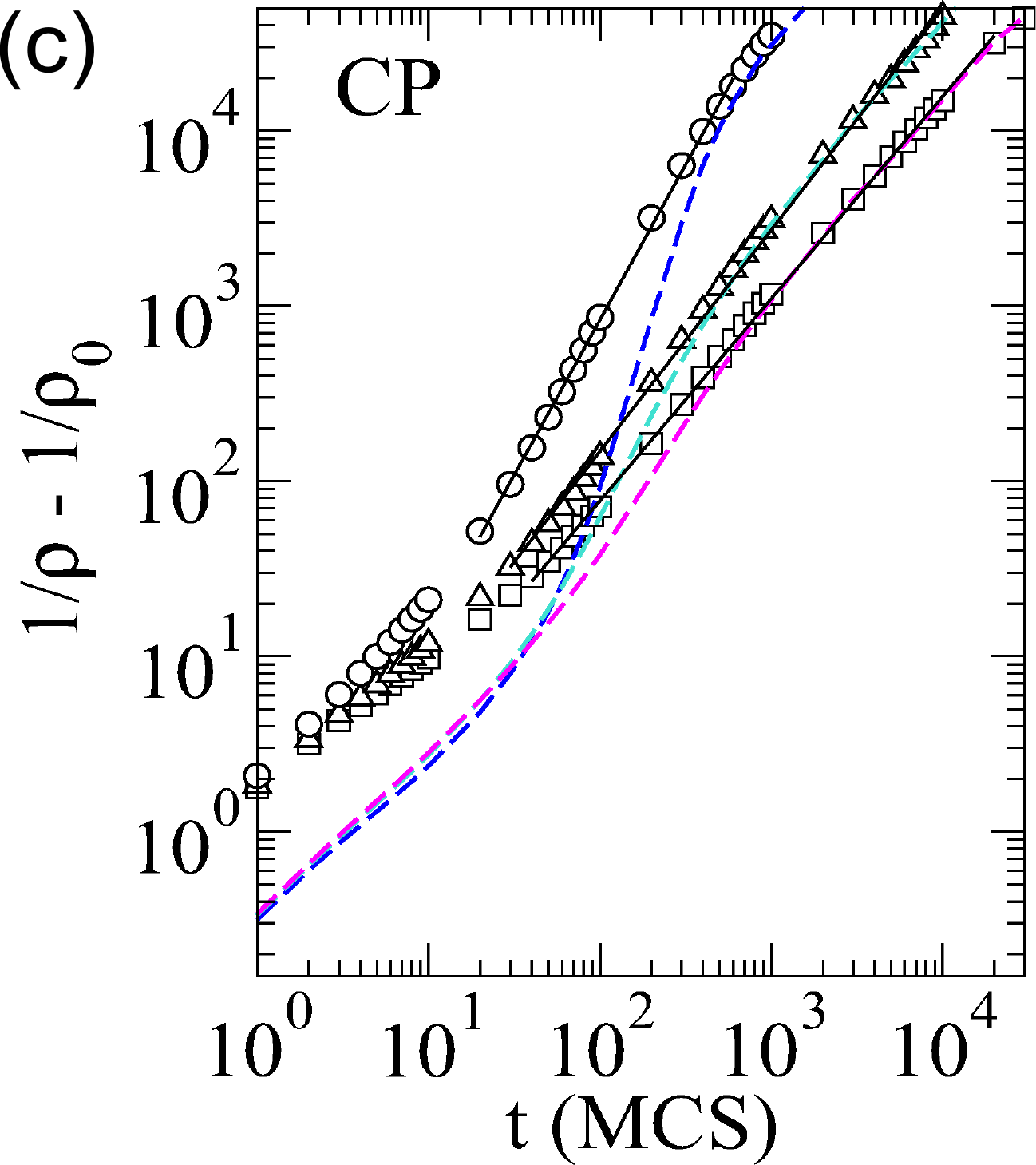}
\hfill
\includegraphics[width=0.24\columnwidth]{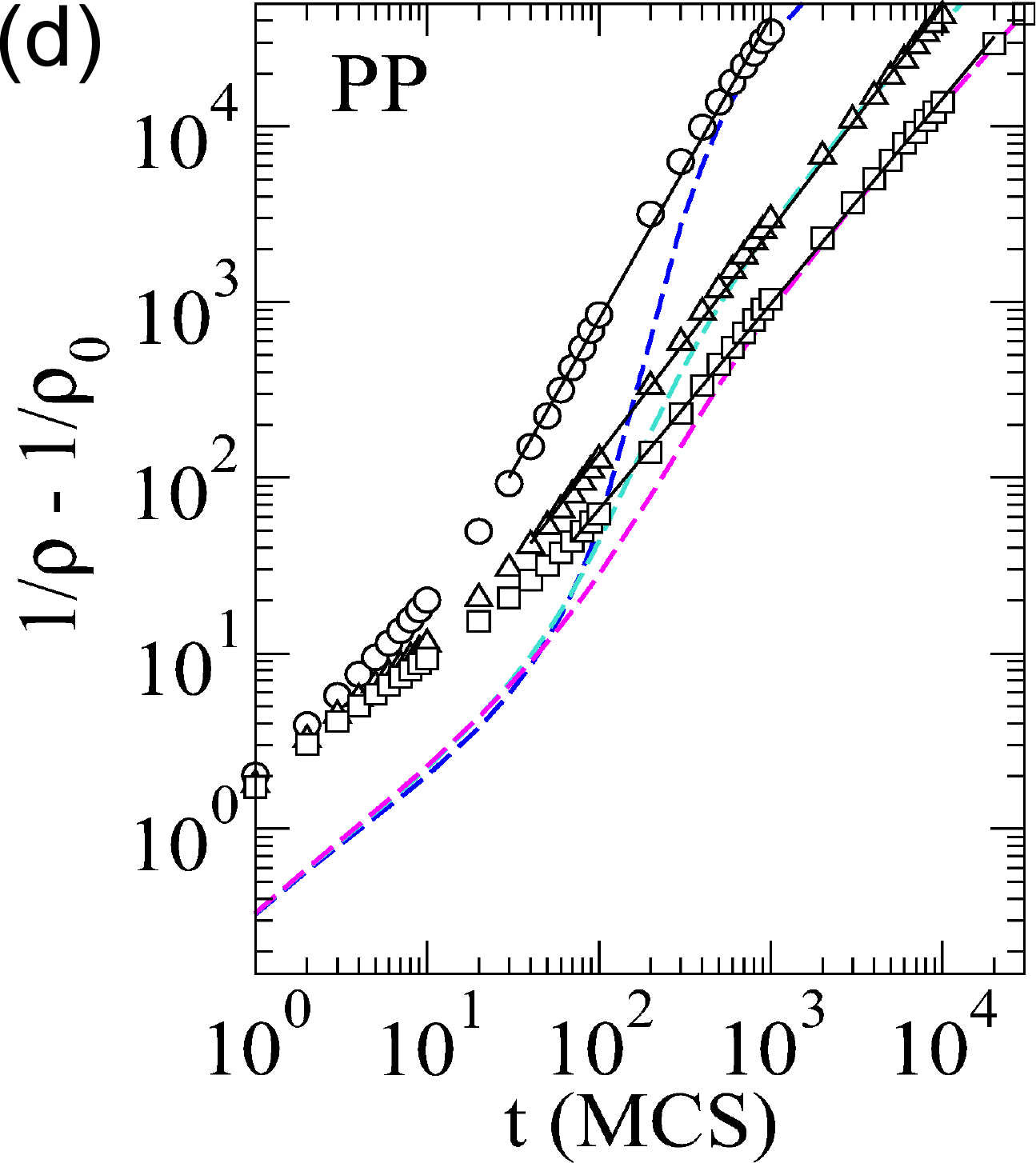}
\caption{(Color online) (a) Illustration of the three different interconnectivity strategies between two scale free networks. The CC strategy
  connects hub nodes from one network to hub nodes from the other, the CP strategy connects hub nodes from one network to
  peripheral nodes in the other, while the PP strategy connects peripheral nodes from one network to peripheral nodes in the
  other. (b)--(d) The reaction progress $1/\rho-1/\rho_{0}$ as a function of time for the three different interconnecting
  strategies of $n$-interconnections, applied on interconnected SFN with $\gamma=2.5,\ 3.0,$ $3.5$, and $q=0.2$. Open symbols 
  show results or the ``well mixed'' case, while dashed lines show results for the ``polarized'' case.  
  The results are averaged over 100 realizations, and the standard errors are smaller than the size of the
  symbols. The continuous lines show the best fit at the asymptotic limit.}
\label{fig:Illustration}
\end{figure*}

This finding, based on computer simulations, showed a rapid acceleration of the process for networks with $\gamma \leq
3.5$. This acceleration was attributed to the existence of hubs, which affect the spatial distribution of particles. The
spatial arrangement of particles at a given time $t$ was studied by the quotient $Q_{\rm AB}(t)$, that measures how well-mixed
the system is by comparing the number of contacts between particles of the same type ($N_{\rm AA}+N_{\rm BB}$) against the number
of contacts between particles of different types ($N_{\rm AB}$)~\cite{Newhouse1988}
\begin{equation}
  Q_{\rm AB}(t)=\frac{N_{\rm AB}(t)}{N_{\rm AA}(t)+N_{\rm BB}(t)}.
  \label{eq:Q}
\end{equation}
Later, this process was treated analytically~\cite{Catanzaro2005,Weber2006}, and subsequent publications tested the effect of the
network generation mechanism on the reaction speed~\cite{Gallos2005a,Gallos2006}.

Here we show how 
the mixing of particles, which determines the reaction rates, depends on the number of interconnecting links and their function,
and we discuss the role of different interconnectivity strategies.

Following Ref.~\cite{Gallos2004}, the SFN used in our study are generated by the standard configuration
model~\cite{Bekessy1972,Bender1978,Bollobas1980,Molloy1998} with $10^5$ nodes, $\gamma=2.5,\ 3.0,$ $3.5$, and $k_{\rm{min}}=1$.  
From the resulting network we extract and use only the largest connected component (LCC). This ensures that the diffusion
process is not biased by isolated network components, but the remaining nodes in our system are less than $10^5$. 
More precisely, the average size of the LCC for our networks was $\sim90000$ nodes for $\gamma=2.5$, $\sim68000$ nodes for $\gamma=3.0$, and $\sim35000$ nodes for $\gamma=3.5$.

In order to create an interconnected system we use two networks generated with the above procedure, and add links between their
nodes. The number $L$ of these interconnecting links is a fraction $q$ of the number of nodes $N$ that are available in each
network, i.e. $L=qN$, and we allow only one interconnecting link per node.
Besides the mere number of interconnecting links, local properties of the interconnected nodes, like their degree $k_{i}$,
can affect the global properties of the system of networks~\cite{Aguirre2014}. In order to test if (and how) degree-degree
correlations between interconnected nodes affect the annihilation reaction's evolution we use three distinct interconnectivity
strategies~\cite{aguirre2013successful}, i) a central to central (CC) strategy that links the $L$ highest degree nodes of the two networks, ii) a peripheral to
peripheral (PP) strategy that links the $L$ lowest degree nodes of the two networks, and iii) a central to peripheral (CP)
strategy that links the $L$ highest degree nodes of one network to the $L$ lowest degree nodes of the second network. An
illustration is shown in Fig.~\ref{fig:Illustration}(a).

Furthermore, interconnecting links can have different functions than the links within each network, so in this work we test two
distinct cases. The first case assumes that the interconnecting links have exactly the same properties and functions as the links
within the networks, and we call them $n$-interconnections. The second case assumes that the interconnecting links have an
``immediate transport'' property, and we call them $t$-interconnections.

In our setting, immediate transport means that if a particle during its diffusive motion in one network lands at a node that is
linked to another network with a $t$-interconnection, the particle is transferred to the interconnected node of the second
network simultaneously. For example, a $t$-interconnection could represent a person that is active in two social networks
and a particle could be a piece of information available to one network. When this information reaches the interconnected person,
it becomes immediately available to the other network\footnote{We should note that in a more realistic setting
  this piece of information should remain on the first network, while at the same time it ``emerges'' on the second one. 
  However, studying such a modified reaction diffusion processes is beyond the scope of the current work.}.

\begin{figure*}[t]
\includegraphics[width=0.49\columnwidth]{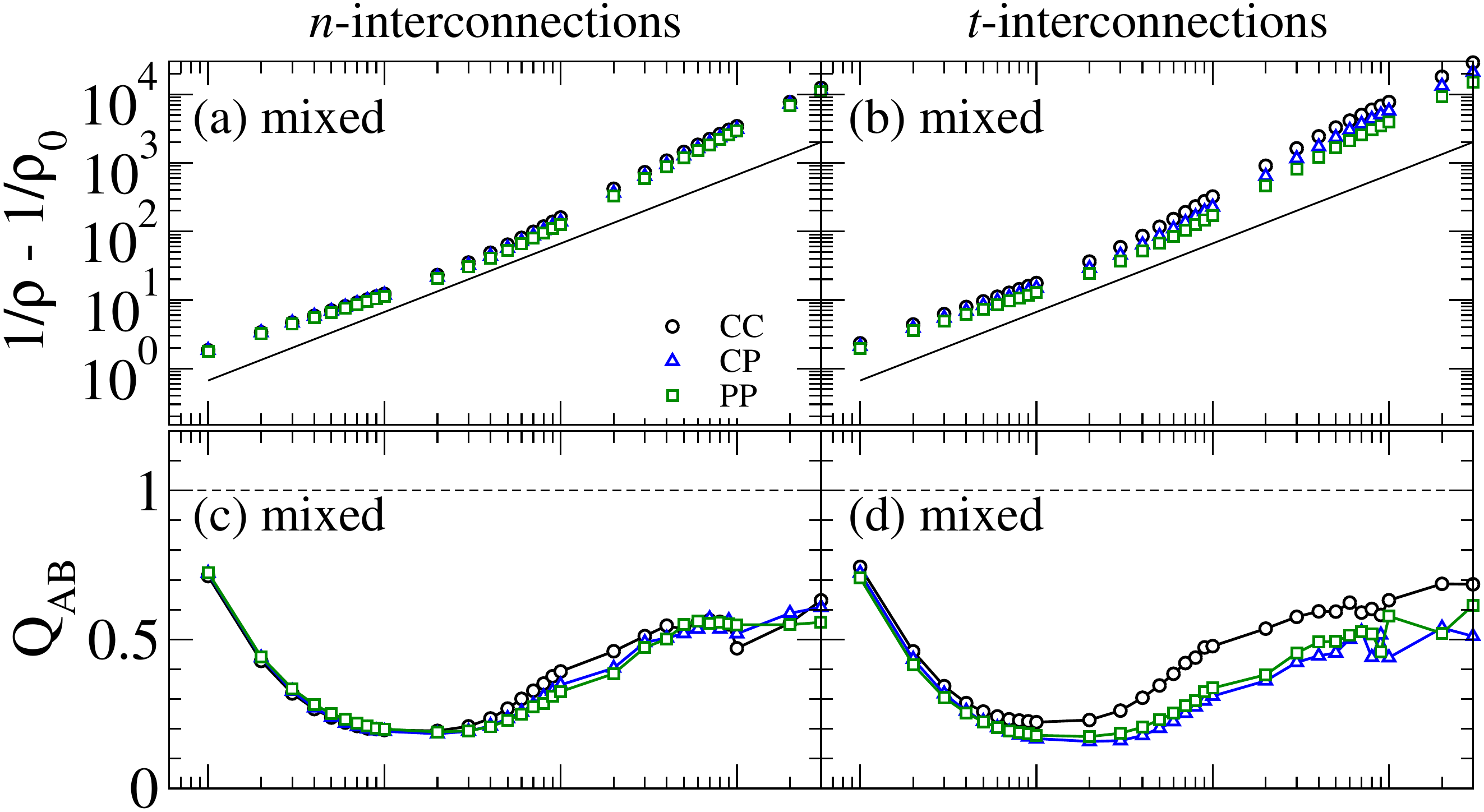}\hfill
\includegraphics[width=0.49\columnwidth]{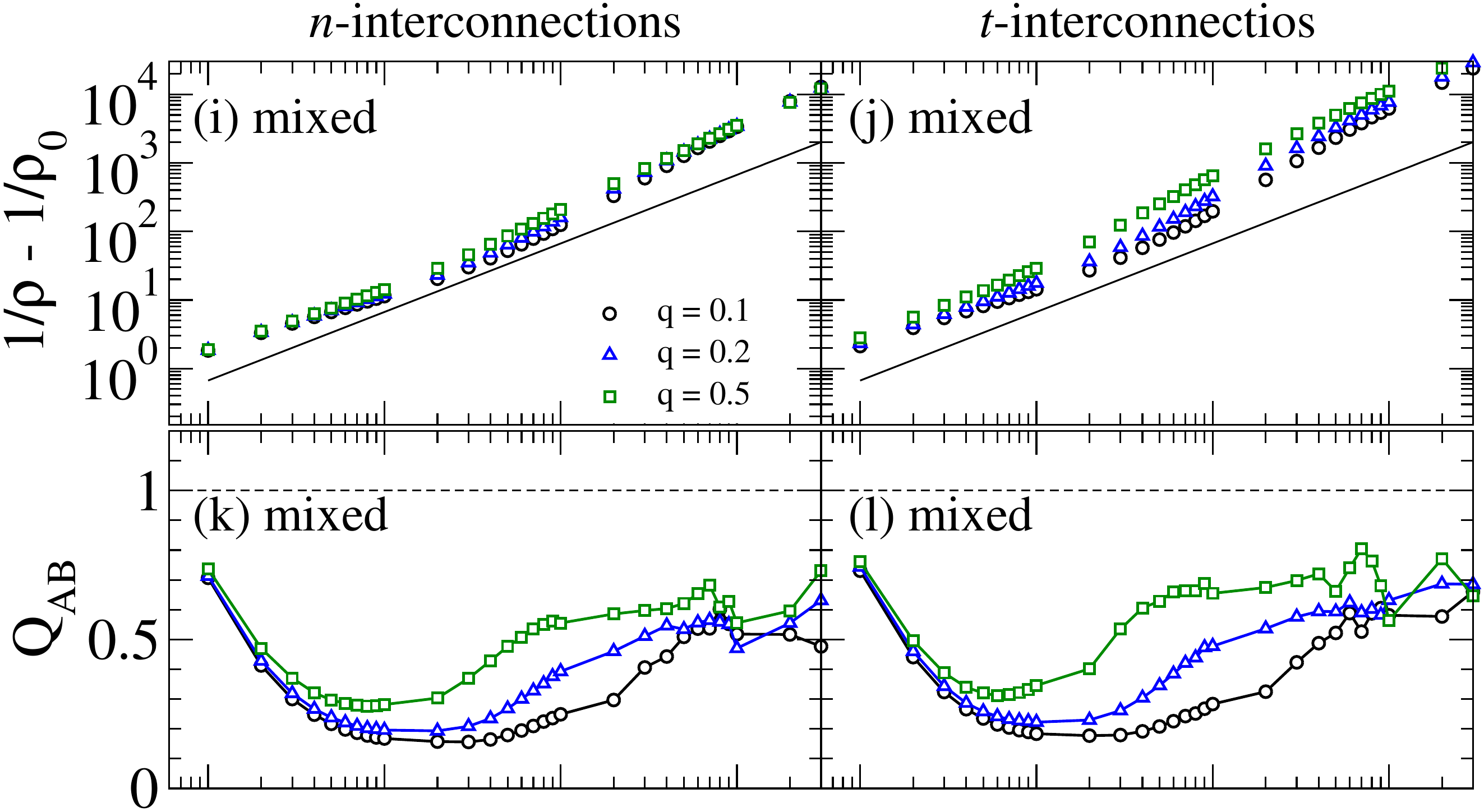}\\
\includegraphics[width=0.49\columnwidth]{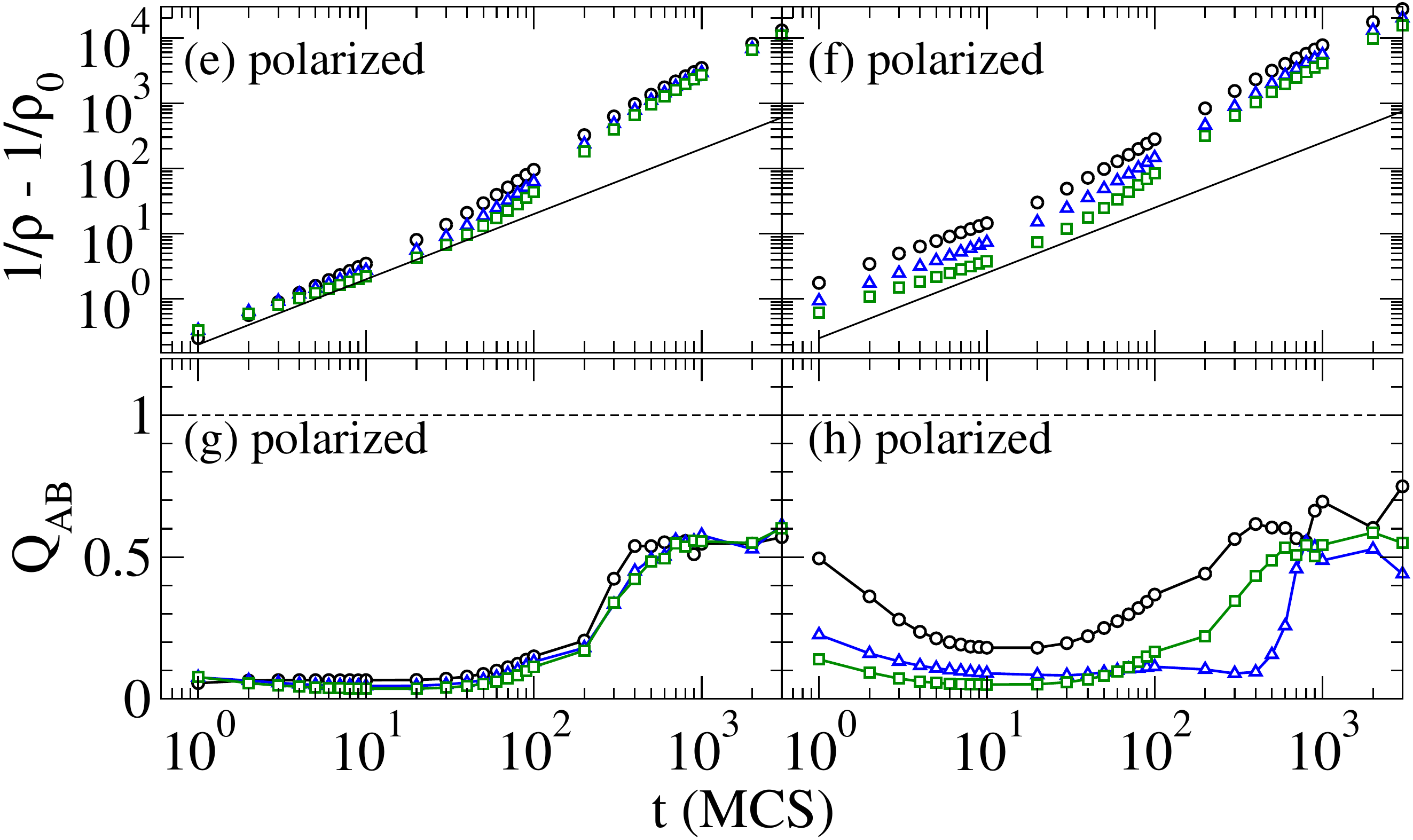}\hfill
\includegraphics[width=0.49\columnwidth]{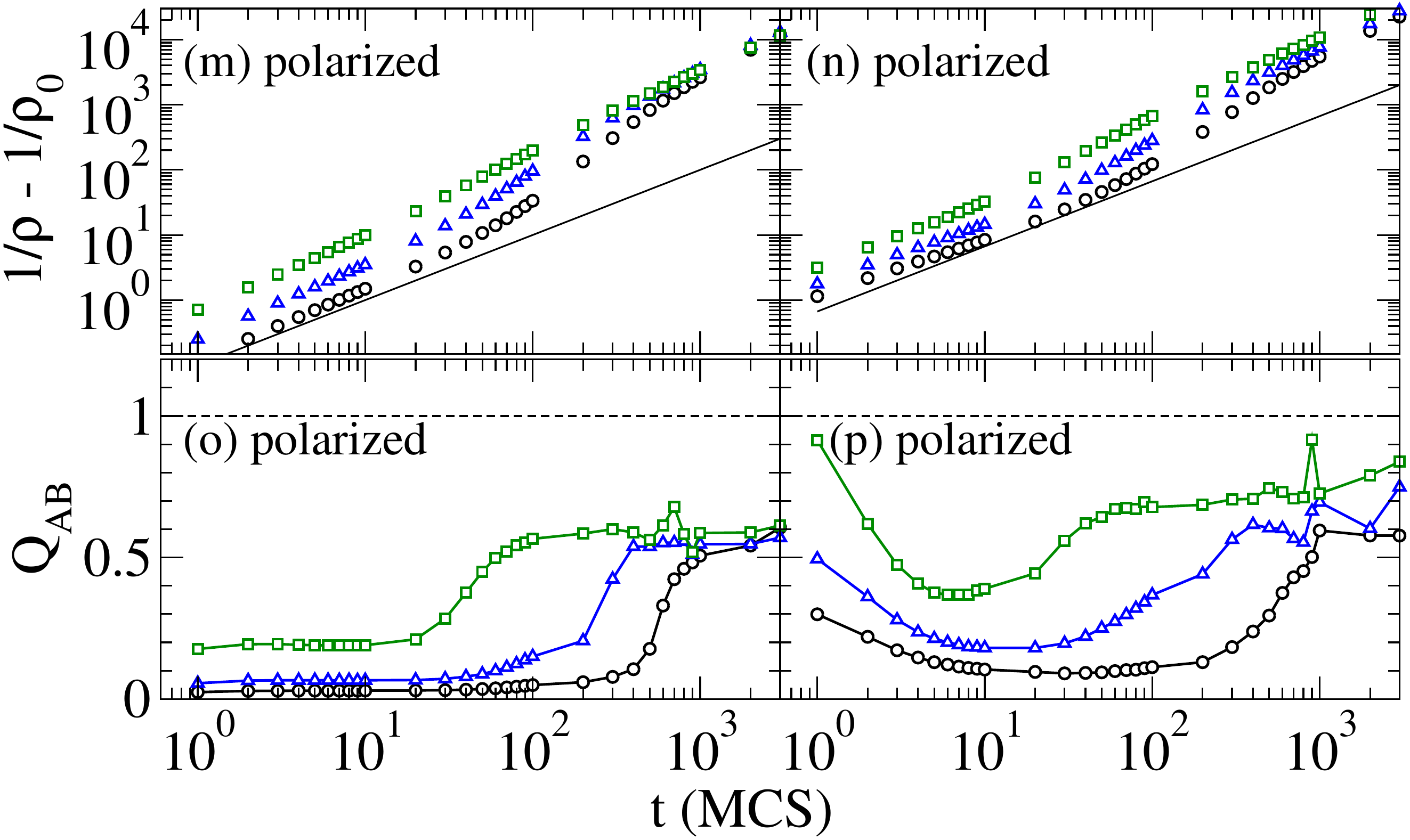}
\caption{(Color online) The reaction progress $1/\rho-1/\rho_{0}$ as a function of time for: 
  the ``well mixed'' case with $n$-interconnections (a) \& (i), and $t$-interconnections (b) \& (j) , and for 
  the ``polarized'' case with $n$-interconnections (e) \& (m), and $t$-interconnections (f) \& (n).
  The ratio $Q_{AB}$ over time for the ``well mixed'' case with $n$-interconnections (c) \& (k), and $t$-interconnections (d) \& (l), and 
  for the ``polarized'' case with $n$-interconnections (g) \& (o), and $t$-interconnections (h) \& (p).  
  All cases are for two coupled SFN with $\gamma = 3.0$. 
  The results are
  averaged over 100 realizations, and the standard errors are smaller than the size of the symbols.  Continuous lines in (a), (b), (e), (f), (i), (j), (m), and (n) 
  show the $1/\rho\sim t$ behavior, and the dashed line at $Q_{AB}=1$ in (c), (d), (g), (h), (k), (l), (o), and (p) corresponds to perfect mixing.
  Left column: The fraction 
  of interconnecting links is $q = 0.2$ arranged according to the
  CC (circle), CP (triangle), and PP (square) strategy. 
  Right column: The fraction 
  of interconnecting links is $q = 0.1$ (circle), $q=0.2$ (triangle), and $q=0.5$ (square), arranged using a CC strategy.
  }
\label{fig:c-Q}
\end{figure*}

We perform Monte Carlo simulations using networks with different exponents $\gamma$, different $q$ values, and different original
configurations.  Our results are averages of 100 realizations per configuration. For every realization
a total number of $M_0 = \rho_0 N$ particles are placed on randomly selected nodes. For simplicity we use equal population of
particles $A$ and $B$, and we set $\rho_0=0.5$, so that in total half of the network's nodes are populated initially.
Furthermore, in order to understand the influence of this initial placement, we test two different configurations. The ``well
mixed'' configuration, which allocates (randomly) half of the $A$ and $B$ particles to one network and half of them to the other,
and the ``polarized'' configuration which places (randomly) all $A$ particles to one network and all $B$ particles to the other
network.

This ``polarized'' configuration, is particularly interesting because it tests the mixing efficiency of the scale free topology in
combination with the different interconnectivity strategies. In diffusion limited reactions, like the one studied here, improving
the mixing efficiency is important as aggregation of particles can significantly slow down the reaction
rates~\cite{Toussaint1983}.

\begin{figure}[t]
\includegraphics[width=0.49\columnwidth]{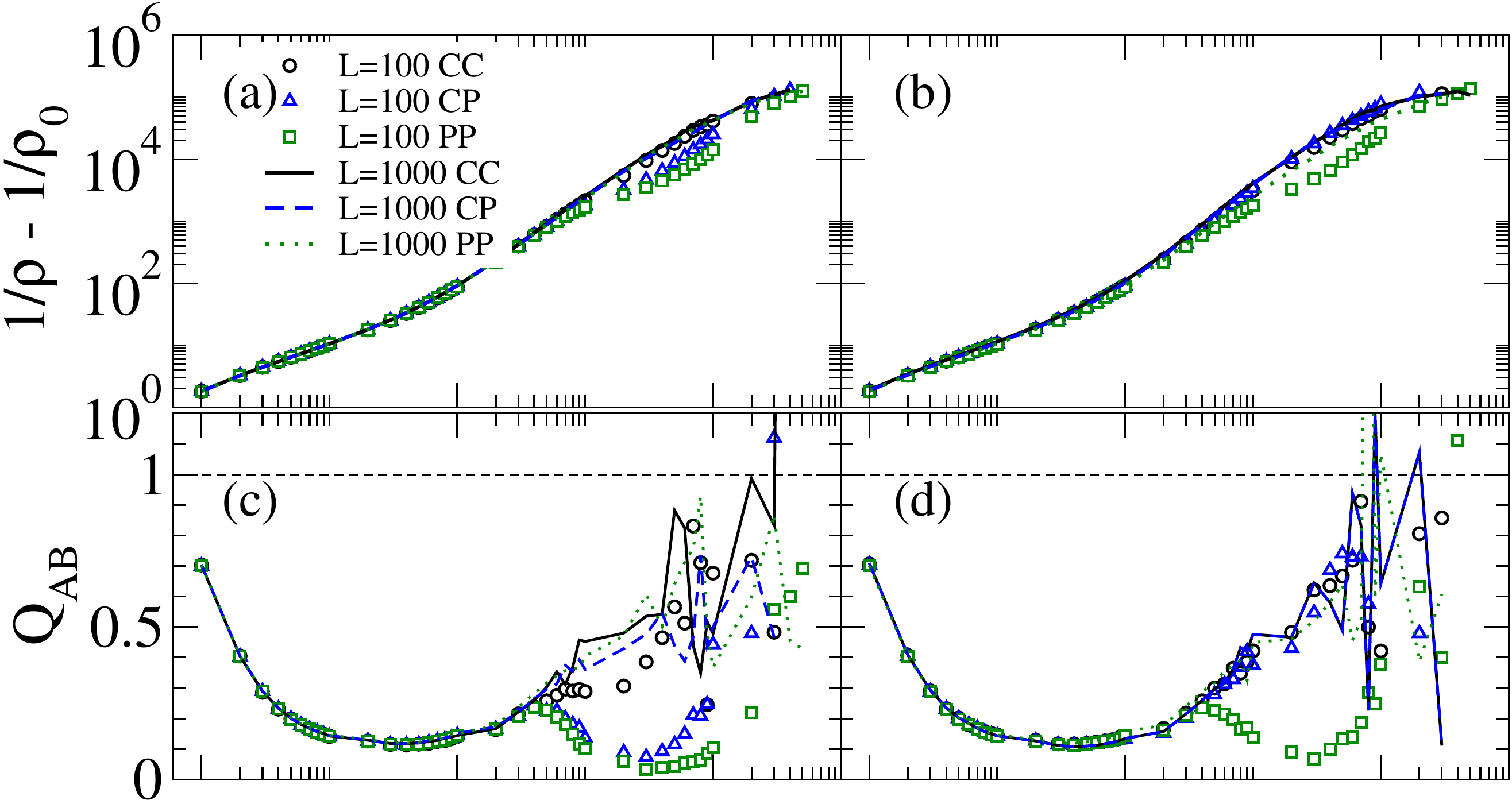}\\
\includegraphics[width=0.49\columnwidth]{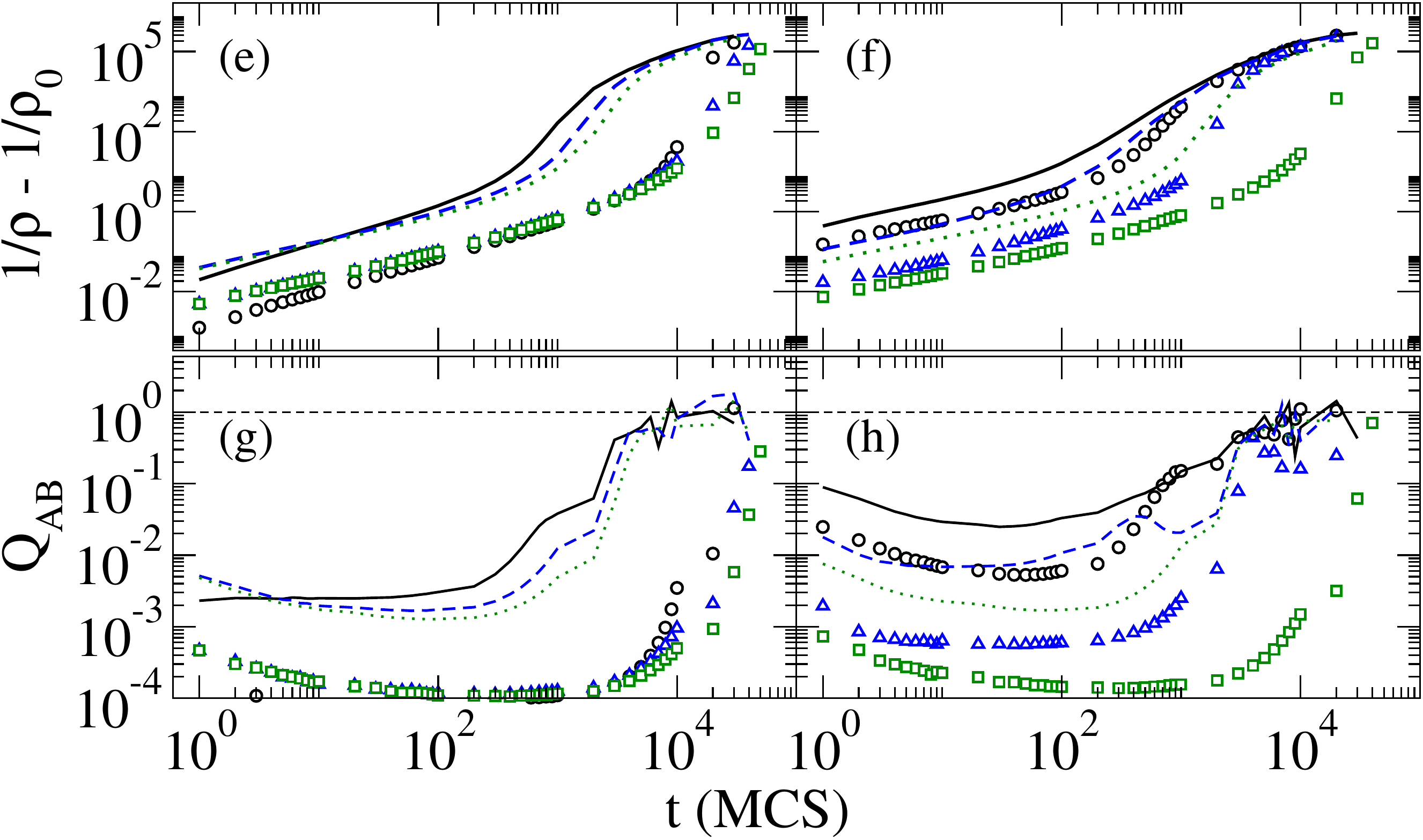}
\caption{(Color online) The reaction progress $1/\rho-1/\rho_{0}$ as a function of time for: 
  the ``well mixed'' case with (a) $n$-interconnections, and (b) $t$-interconnections, and for 
  the ``polarized'' case with (e) $n$-interconnections, and (f) $t$-interconnections.
  The ratio $Q_{AB}$ over time for the ``well mixed'' case with (c) $n$-interconnections, and (d) $t$-interconnections, and 
  for the ``polarized'' case with (g) $n$-interconnections, and (h) $t$-interconnections.  
  All cases are for two coupled SFN with $\gamma = 3.0$, with a fixed number $L=100$ (symbols) or $L=1000$ (lines) of interconnecting links.
  The results are
  averaged over 100 realizations, and the standard errors are smaller than the size of the symbols.
  }
\label{fig:c-Q-L}
\end{figure}

For our simulations we select an occupied node at random, together with one of its neighboring nodes. If the neighboring node is
empty, the particle moves and occupies a new position (diffusion phase). If the neighboring node is already occupied (reaction
phase), then the particles annihilate if they are of different types, while if they are of the same type nothing happens and the
chosen particle remains at its original position. This procedure is repeated by continuously selecting, moving and (possibly)
annihilating particles, until there are no particles left in the system.  Due to the annihilation process the total number of
particles $M(t)$ is reduced with time. Thus, the time (in Monte Carlo Steps -- MCS) required to update the system's state is
advanced inversely proportional to the current number of particles, by 1/M(t).

Here we would like to stress that additional care is taken for the treatment of interconnecting links. For the case of
$n$-interconnections, the interconnecting link behaves like all other links, and the particle has the same probability to ``jump''
to the other network through the interconnecting link, as it has to follow any other link to neighboring nodes in the same
network. In contrast, a $t$-interconnection is not considered
during the diffusing motion of the particle, but only at the end when the particle has moved already. In this case, if the
particle moves to a node that is $t$-interconnected then it is immediately transported to the other network. Also, after
generating the initial configuration --and only for the case of $t$-interconnections-- if there are interconnected nodes occupied
by particles of different types, then these particles are annihilated immediately.

As shown in Figs.~\ref{fig:Illustration}(b)-(c), for the case of $n$-interconnections and a well mixed system, our results are
comparable to Ref.~\cite{Gallos2004} for all different strategies. More precisely, the \emph{absence of kinetic effects} seems to
depend more on the exponent $\gamma$ than on the strategy we use to interconnect the nodes.
For example, for the CC strategy we find $f=1.7\pm 0.02$ for $\gamma=2.5$, $f=1.25\pm 0.01$ for $\gamma=3.0$, and $f=1.14\pm 0.01$
for $\gamma=3.5$. However, if we focus on a specific exponent, e.g. $\gamma=3.0$, then we find $f=1.24\pm 0.01$ for CC,
$f=1.26\pm 0.01$ for CP, and $f=1.28\pm 0.01$ for PP.
This shows that an interconnected system of two SFN with $n$-interconnections behaves like one large SFN for a reaction-diffusion
process.
Also, the same asymptotic behavior holds even when the system starts from a completely polarized configuration. Of course in this case
initially the reaction rates are low, as expected but, as time advances the system becomes better mixed and the
reaction becomes faster. This essentially means that a scale free topology alone contributes to the
mixing of particles and enhances desegregation.

This is better visible in Fig.~\ref{fig:c-Q}(a), where the mean-field predicted limiting case $1/\rho\sim t$ is
included. When viewed together with Fig.~\ref{fig:c-Q}(c), it becomes clear that the different strategies do not really
affect the mixing of particles, and as a consequence the reaction rates.  However, while even with the use of $t$-interconnections
we obtain similar findings with respect to the reaction rates, as shown in Fig.~\ref{fig:c-Q}(b), the CC strategy is now
more effective in particle mixing, at least in intermediate time scales, as shown in Fig.~\ref{fig:c-Q}(d).

This attribute of the CC strategy is more pronounced when we start with an extremely polarized case. 
More precisely, while the overall mixing and reaction rates are similar in the case of
$n$-interconnections (see Figs.~\ref{fig:c-Q}(e)\&\ref{fig:c-Q}(g)), the better mixing achieved by the CC strategy is clear in the
presence of $t$-interconnections. In this case, as shown in Figs.~\ref{fig:c-Q}(f)\&\ref{fig:c-Q}(h), the $Q_{\rm AB}$ values for CC are
higher at all times, and the concentration of particles decreases much faster. However, at longer times the mixing achieved by
the other strategies improves, while at the same time the number of particles decreases. This leads to a crossover point around
$t=11$ MCS where the reaction rates increase rapidly, as shown by the values of $f$ (i.e. $f=1.44 \pm 0.02$ for CC, $f=1.57 \pm
0.02$ for CP, and $f=1.69 \pm 0.05$ for PP ~\footnote{Please note that we are interested in the limit behavior for large $t$, therefore, the fit is performed in the area after the crossover point.}) that converge to a similar asymptotic behavior for all three strategies
(Fig.~\ref{fig:c-Q}(f)).

Similar observations are made for different $q$ values as shown in the right panel of Fig.~\ref{fig:c-Q}. In this case
it is easier to see that higher $q$ values result to better mixing of particles, as expected since the system becomes better
connected. In addition, the mixing is enhanced even more by the presence of $t$-interconnections, as shown by comparing
Fig.~\ref{fig:c-Q}(k) to Fig.~\ref{fig:c-Q}(l), and Fig.~\ref{fig:c-Q}(o) to Fig.~\ref{fig:c-Q}(p). However,
the same conclusions are true with respect to increasing reaction rates at longer times, which allows the convergence
to the same asymptotic behavior.

On the other hand, even $q=0.1$ for large SFN  leads to large number of interconnections, while at the same time very few nodes are hubs.
Thus, the CC strategy may not always connect central to central nodes, and its potential influence may be masked.
For this reason we repeated our analysis using smaller numbers of interconnecting links, namely $L=1000$ and $L=100$, and in Fig.~\ref{fig:c-Q-L} we show our results.
Even though  for $L=1000$ the overall reaction rate is similar for all strategies, especially for the well mixed system, some deviations are visible for $L=100$ where the CC/PP strategy leads to better/worse mixing.
Therefore, the influence of the different strategies becomes visible in the presence of few interconnections, where the CC strategy clearly leads to better mixing.
But, when the number of interconnecting links is large enough, then the number and function of interconnections resume the most important role in driving the diffusion process.

Summarizing, we studied the annihilation reaction $A+B\rightarrow \emptyset$ in interconnected SFN, when different
interconnectivity strategies are used, and when the interconnecting links have different functionality than the normal links. We
find that the system of networks exhibit rapid reaction rates, in line with a previous observation about single
SFN~\cite{Gallos2004} -- which is different from what was observed in other topologies, like lattices and fractals. In addition,
we showed that the CC strategy is better for the mixing of particles when there are few interconnections, but for larger number of interconnections the function and the number of interconnecting links
plays the most important role. We thus identified ways that can be used to suppress the
segregation phenomenon and enhance the diffusion of particles. Besides their relevance to reaction kinetics, our findings could be
applied to model propagation of conflicting information in social networks, and identify ways to reduce polarization.

\paragraph*{Acknowledgements.}
The author acknowledges support by the EU-FET project MULTIPLEX 317532.

\clearpage
\newpage
\bibliographystyle{apsrev4-1}
\bibliography{Interconnected-Annihilation}
\end{document}